\documentclass{iaus}
\usepackage{graphicx,natbib,amsmath}
\newcommand{\f}[1]{\smash{\hat #1}}                
\newcommand{\V}[1]{\boldsymbol{#1}}                
\newcommand{\wA}{\omega_\mathrm{A}}

\title[MHD Turbulence in Accretion Disk Boundary Layers]
{On Magnetohydrodynamic Turbulence \\
  and Angular Momentum Transport \\
  in Accretion Disk Boundary Layers}

\author[C.-K. Chan \& M.~E. Pessah]
{Chi-kwan Chan$^1$ \and Martin~E. Pessah$^2$}

\affiliation{$^1$NORDITA, Roslagstullsbacken 23, 106 91 Stockholm, Sweden \\
  email: {\tt ckch@nordita.org} \\
[\affilskip] $^2$Niels Bohr International Academy, Niels Bohr Institute\\
  Blegdamsvej 17, 2100 Copenhagen \O, Denmark \\
  email: {\tt mpessah@nbi.dk}}

\pubyear{2012}
\volume{294}
\jname{Solar and Astrophysical Dynamos and Magnetic Activity}
\editors{A.G. Kosovichev, E.M. de Gouveia Dal Pino, \& Y.Yan, eds.}

\begin{document}

\maketitle

\begin{abstract}
  The physical modeling of the accretion disk boundary layer, the
  region where the disk meets the surface of the accreting star,
  usually relies on the assumption that angular momentum transport is
  opposite to the radial angular frequency gradient of the disk.
  The standard model for turbulent shear viscosity, widely adopted in
  astrophysics, satisfies this assumption by construction.
  However, this behavior is not supported by numerical simulations of
  turbulent magnetohydrodynamic (MHD) accretion disks, which show that
  angular momentum transport driven by the magnetorotational
  instability is inefficient in this inner disk region.
  I will discuss the results of a recent study on the generation of
  hydromagnetic stresses and energy density in the boundary layer
  around a weakly magnetized star.
  Our findings suggest that although magnetic energy density can be
  significantly amplified in this region, angular momentum transport
  is rather inefficient.
  This seems consistent with the results obtained in numerical
  simulations and suggests that the detailed structure of turbulent
  MHD boundary layers could differ appreciably from those derived
  within the standard framework of turbulent shear viscosity.

\end{abstract}

\firstsection

\section{Introduction}

For an accretion disk to merge smoothly to the surface of the central
accreting star, its angular velocity $\Omega(r)$ must decrease inwards
in the inner disk to match the sub-Keplerian stellar spin rate
$\Omega_\star$ \citep{2002apa..book.....F, 2009apsf.book.....H,
  2010apf..book.....A}.
It is easy to derive from the standard accretion disk model that half
of the accretion energy is released in this so called accretion disk
boundary layer (ADBL).
Understanding the detailed physics within this thin layer, therefore,
is of great astrophysical interest \citep{2004ApJ...610..977P,
  2009ApJ...702.1536B, 2010AstL...36..848I}.

According to standard models, the structure of ADBL depends entirely
on turbulent angular momentum transport.
On the one hand, standard models implicitly assume that the
\emph{turbulent stress} is linearly proportional to the local shear
\citep{1973A&A....24..337S}.
On the other hand, it is widely accepted that the magnetorotational
instability (MRI) is responsible to drive turbulence in the accretion
disk \citep{1991ApJ...376..214B, 1998RvMP...70....1B}.
Given that the shear rate, $q \equiv - d\ln\Omega/d\ln r$, is negative
in the boundary layer, the MRI is inactive and the standard ADBL model
breaks down.

To understand the behavior of the ADBL and derive a self-consistent
model, we study the local evolution of magnetohydrodynamic (MHD) waves
when they are stable to the MRI.
We explain how shearing MHD waves can generate strong toroidal
magnetic fields without leading to efficient angular momentum
transport.

\section{Local Model for MHD Accretion Disk Boundary Layers}

We employ the standard shearing box approximation
\citep{1995ApJ...440..742H} and follow the notation in
\citet{2012ApJ...751...48P} to denote the velocity and magnetic
fluctuations by $\V{u}$ and $\V{b}$, respectively.
By assuming the flow is incompressible and considering only a single
(shearing) Fourier mode at wavenumber $\V{k} \equiv (k_x, k_y, k_z)$,
all the non-linear terms in the ideal MHD equations vanish identically
\citep{1994ApJ...432..213G}.

To isolate the interesting non-MRI dynamics, we focus only on
$z$-independent perturbation.
That is, we set $k_z \equiv 0$ .
The non-trivial MHD can be summarized in the following second-order
ordinary differential equation \citep[see][for
  derivation]{2012ApJ...751...48P}:
\begin{align}
  \frac{d\f{b}_x}{d\tau^2} + \Gamma(\tau)\frac{d\f{b}_x}{d\tau} +
  \omega^2 \f{b}_x = 0.
  \label{eq:2nd}
\end{align}
In the above equation, the independent variable $\tau \equiv
k_x(t)/k_y = q\,\Omega_0 t$ is the dimensionless time, where
$\Omega_0$ is the corotating frequency of the shearing box; the
dependent variable $\f{b}_x$ is the Fourier coefficient of $b_x$ at
$\V{k}$.
The symbol $\omega \equiv \wA/q\Omega_0$ is the dimensionless form of
the (constant) Alfv\'en frequency $\wA \equiv \V{B}_0\cdot\V{k}$,
where $\V{B}_0$ is the background magnetic field.
Finally, we use the shorthand $\Gamma(\tau) \equiv 2\tau/(\tau^2 +
1)$.
Other components of the fluctuations can be easily derive from
$\f{b}_x$ by
\begin{align}
  \f{u}_x = -\frac{i}{\omega} \frac{d\f{b}_x}{d\tau}
\end{align}
and the divergence-less conditions $\nabla\cdot\V{u} =
\nabla\cdot\V{b} = 0$.

\begin{figure}
  \center\includegraphics[trim=16 8 8 16]{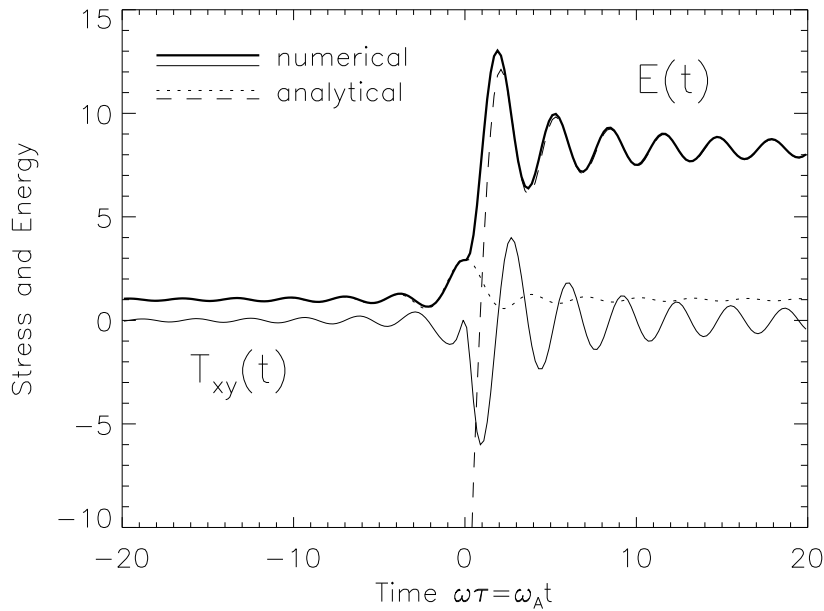}
  \caption{The thick and thin solid lines correspond, respectively, to
    the numerical total energy $E(t)$ and total stress $T_{xy}(t)$.
    The dotted and dashed lines show the analytic approximation for
    the energy, equation~(\ref{eq:E}).}
  \label{fig:se}
\end{figure}

Equation~(\ref{eq:2nd}) does not have an analytic solution.
Nevertheless, it reduces to a Bessel equation in the limit $\tau
\rightarrow \pm\infty$.
By keeping up to the first order terms in $1/\omega\,\tau$, we can
derive the (mean) total stress $\smash{T_{xy}} \equiv \langle
\smash{u_x u_y} - \smash{b_x b_y}\rangle$ and the (mean) energy
density $E\equiv \langle \smash{u^2 + b^2} \rangle/2$, where
$\langle\ \cdot\ \rangle$ stands for the spatial average:
\begin{align}
  \!T_{xy} \!\approx\!-\frac{2}{\omega^2\tau}
  &\Big[(|S|^2 - |C|^2)\cos(2\omega\,\tau)
      + ( S^*C +  SC^*)\sin(2\omega\,\tau)\Big],
  \label{eq:Txy}\\
  E \!\approx\! -\smash{\frac{1}{\omega^3\tau}}
  &\Big[(|S|^2 - |C|^2)\sin(2\omega\,\tau)
      - ( S^*C +  SC^*)\cos(2\omega\,\tau)\Big]
      + \frac{|S|^2 - |C|^2}{\omega^2}.
  \label{eq:E}
\end{align}
The complex integration constants $S$ and $C$ are determined by the
initial conditions; and the star symbol denotes complex conjugation.
Note that the energy balance equation $dE/dt = q \Omega_0 T_{xy}$ is
satisfied up to first order.

We plot the numerical solutions for the energy $E(t)$ and the stress
$T_{xy}(t)$ in Figure~\ref{fig:se} by thick and thin solid lines,
respectively.
The dotted and dashed lines show our analytic approximation with two
different sets of integration constants.
Note that the energy increases significantly around $\tau = 0$
although the stress keeps oscillating around zero.
The physical meaning is that the shearing MHD waves can extract energy
from the background without increasing the time-averaged stress.
Hence, the dissipation of magnetic energy in the ADBL is inefficient.

\begin{figure}
  \center\includegraphics[trim=16 8 8 16]{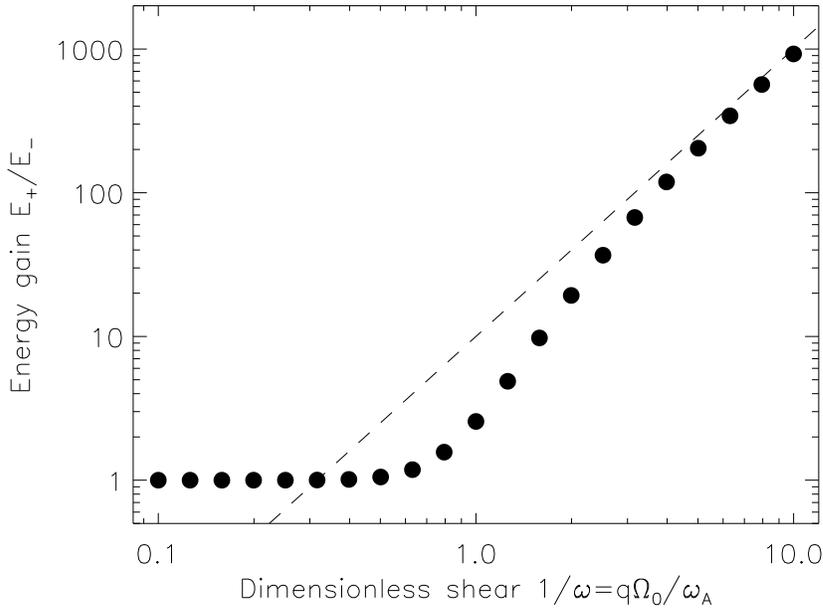}
  \caption{The filled circles represent the value of the amplification
    factor $E_+/E_-$ for different values of the shear.
    The dashed line shows the function $10/\omega^2 \equiv
    10(q\Omega_0/\wA)^2$, which is in good agreement with the
    numerical results for strong shear.
    This asymptotic behavior is insensitive of the initial conditions
    as long as $C_- \ne 0$.
    See \citet{2012ApJ...751...48P} for details.}
  \label{fig:rq}
\end{figure}

Let $E_\pm \equiv \lim_{\tau\rightarrow\pm\infty}E(\tau)$.
Despite the fact that we cannot analytically solve for the
amplification factor $E_+/E_-$, we can still draw important
conclusions from the properties of the numerical solutions.
We plot the amplification factor as a function of $1/\omega$ in
Figure~\ref{fig:rq}.
In the limit of weak shear, the pure Alfv\'en waves cannot lead to net
energy gain.
Nevertheless, for strong shear $q \gg \wA/\Omega_0$, or $1/\omega \gg
1$, the amplification factor follows a power law,
\begin{align}
  \frac{E_+}{E_-} \approx \frac{10}{\omega^2} =
  10\left(\frac{q\Omega_0}{\wA}\right)^2.
\end{align}
This is remarkable because the amplification is unbounded without a
linear instability.

\section{Conclusions}

Understanding the relationship between the stress and the shear rate
is essential in modeling the ADBL.
In this proceeding, we point out that the standard turbulent accretion
disk model is inapplicable in the boundary layer because the MRI is
inactive there.
We provide the key steps to solve stable shearing MHD waves, which
describes the physics better.
Although the energy of the waves can be significantly amplified, the
time-averaged stress remains zero.
Our findings agree with global MHD simulations of accretion disks
performed by \citet{2002MNRAS.330..895A, 2002ApJ...571..413S}.

Our results suggest that the ADBL will be strongly magnetized.
However, the magnetic fields will not be able to transport angular
momentum nor dissipate to heat up the plasma.
One direct consequence is that the angular momentum transport from an
accretion disk to a central accreting star, as predicted by the
standard ADBL models, is suppressed.
Some other physical mechanism is therefore needed to spin up weakly
magnetized stars.

\bibliography{ms}

\end{document}